\documentclass[aps,prl,twocolumn]{revtex4}
\usepackage{color,amsmath}
\newcommand{\br}{\mathbf{r}}
\begin{document}

\title{Diffusive density profiles in a cold-atom expansion experiment}

\author{Cord A. M\"uller$^{1,2}$ and Boris Shapiro$^3$}

\affiliation{$^1$ Fachbereich Physik, Universit\"at Konstanz, 78457
 Konstanz, Germany \\
$^2$ Centre for Quantum Technologies, National University of
Singapore, Singapore 117543, Singapore\\
$^3$Department of Physics, Technion - Israel Institute of Technology, Haifa 32000, Israel}

\maketitle

In a recent experiment \cite{deMarco}, the expansion of
non-interacting ultracold fermions was studied in a random speckle
potential, and the observed density profiles were interpreted based on 3D Anderson localization.
The purpose of this Comment is to demonstrate that
slow diffusion of particles with a
broad energy distribution and an energy-dependent
diffusion coefficient leads to density
profiles that agree with measured data of
\cite{deMarco}, but not with the behavior expected for 3D Anderson
localization.

Consider a non degenerate  Fermi gas
prepared in a trap with the  semiclassical position-energy distribution $F(\br,
E)$. At time $t=0$ the trap is
switched off and atoms start to spread in the random potential.
Assuming diffusive spreading, the ensemble averaged density is
\begin{eqnarray}
\overline{n}(\textbf{r},t) = \int d\textbf{R}\int dE
P_{E}(\textbf{r},\textbf{R},t) F(\textbf{R}, E),
\label{density}
\end{eqnarray}
where the $d$-dimensional diffusion kernel
$P_{E}(\textbf{r},\textbf{R},t)=\left(4\pi
D_{E}t\right)^{-d/2}
\exp\left(-|\br-\mathbf{R}|^{2}/4D_{E}t\right)
%\label{kernel}
$ 
propagates a particle with kinetic energy $E$ and diffusion constant $D_E$ from
initial position $\textbf{R}$ to final position $\br$ in time
$t$.
In the long time limit, when the atoms have spread over distances much
larger than the initial cloud, the starting point $\textbf{R}$  in
the kernel can be set to zero. The
remaining integral over $\textbf{R}$ yields the atomic energy
distribution, $\int d\textbf{R} F(\textbf{R},
E) = C\nu(E)\exp(-\beta E)$, where $C$ is a
normalization constant, $\nu(E)$ is the density of states,
and  $\beta$ measures the distribution width, generally due to
temperature as well as disorder. Eq.~(\ref{density}) then reduces to
\begin{eqnarray}
\overline{n}(\textbf{r},t) = C\int dE
\frac{\nu(E)}{(4\pi D_E t)^{d/2}}
\exp(-\frac{r^2}{4D_E t} - \beta E). \label{cloud}
\end{eqnarray}
Thus, the density profile of a diffusing atomic cloud consists of a rather sharp
   central region, due to slowly diffusing particles piling up close to the center, followed by a
   (stretched)-exponential tail (see \cite{rev} for an analytically
   tractable example). Indeed,
the tails of the profile (\ref{cloud}) are determined by a saddle point of the
expression in the exponent.
When the energy dependence of
the diffusion coefficient is a pure power
law $D_{E} = \hbar E^a / m E_0^a$ (with $E_0$ an
appropriate energy scale), this yields to leading order
\begin{equation}
\overline{n}(\textbf{r},t) \sim \exp\{-|r/s(t)|^{2/(a+1)}\},
\label{tail}
\end{equation}
valid for $r^2\gg s(t)^2 =s_a^2\hbar t/[m (\beta E_0)^{a}]$ with
$s_a$ of order unity.
This is a stretched exponential with exponent $2/(a+1)$
depending on dimensionality and type of disorder.

The experiment \cite{deMarco} uses a very strong and smooth random
potential, with an axial correlation longer than the de
Broglie wavelength of most atoms, such that $D_{E} \propto E
^{5/2}$, over a sizable range of energies down to the rms potential
strength  \cite{rev}.
The density tail is then the
stretched exponential $|\log\overline{n}(\textbf{r},t)| \propto
r^{4/7}$.
It is important to note that eq.~\eqref{density} describes atoms
in an intermediate energy interval. Atoms with higher energies
appear ballistic in the finite field of observation, whereas atoms
with lower energies should localize. The value 4/7 for the exponent is just an estimate,
although it lies
in a reasonable range judging from the supplemental notes of \cite{deMarco}.
Our main point
here is that the density profiles of \cite{deMarco} are
characteristic of a slowly diffusive component and are incompatible with the localization scenario  which
should  lead to a power law tail (see below).

Even for the strongest disorder the data in Fig.~3(d) of
\cite{deMarco}
suggest that the rms radius of the cloud keeps growing
instead of saturating 
on the observed time scale. Although it is
difficult to determine an accurate slope, 
the residual dynamics could correspond to a diffusion coefficient of
$40\,{\mu\text{m}}^2/$ms, which is much larger than the ``quantum of
diffusion'' $\hbar /m\approx 1.5\,\mu \text{m}^2/$ms of a potassium
atom \cite{rev}. Thus, this data is compatible with diffusion at energies above the mobility edge.
For some of  the data in Fig.~3(d), occasionally, the size
of the atomic cloud even appears to shrink. Systematic
contraction is of course incompatible both with diffusion as well as localization, and should be analyzed with great care in view of the particle losses that limit the
lifetime of the trapped gas---a task that is beyond the scope of the present
Comment.

Lastly, there is an additional argument against a localization scenario. The cloud spreads to a rather large distance, of the
order of $1\,$mm. If 
atoms are 
localized on such a large scale, 
they must originate from the critical energy interval next to
the mobility edge, where the localization length diverges. But
the resulting stationary density tail is well known to be a power law,
$\overline{n} \sim  r^{-\left(3 +(1/\nu)\right)}$,
(here $\nu$ is the localization length exponent) instead of an exponential   \cite{rev}.
While this result was derived for a short-range, statistically
isotropic potential, such
power-law tails are expected to occur for any generic random
potential, including anisotropic ones. 
Indeed, their existence is based solely on the fact that the localization length 
diverges as a power law when the
mobility edge is approached from 
below, which is the generic behavior at the Anderson transition.

%%%%%%%%%%%%%%%%%%%%%%%%%%%%%%%%%%%%%%%%%%%%%%%%%%%%%%%%%%%%%%%%%%%%%%%%%%%%%%%%%%%%%%%

%%%%%%%%%%%%%%%%%%%%%%%%%%%%%%%%%%%%%%%%%%%%%%%%%%%%%%%%%%%%%%%%%
\end{document}